\documentclass[conference,a4paper]{APSIPA2026}
\IEEEoverridecommandlockouts
\usepackage{cite}
\usepackage{amsmath,amssymb,amsfonts}
\usepackage{graphicx}
\usepackage{textcomp}
\usepackage{xcolor}
\usepackage{url}
\usepackage{algorithm}
\usepackage{algpseudocode}
\usepackage{bm}
\usepackage{booktabs}

\usepackage{geometry}
\usepackage{fancyhdr}

\def\BibTeX{{\rm B\kern-.05em{\sc i\kern-.025em b}\kern-.08em
    T\kern-.1667em\lower.7ex\hbox{E}\kern-.125emX}}
\begin{document}

\title{CraBERT: Efficient Phoneme Encoder Pre-Training via Cascade Fusion of Subword Representations \\ for Text-to-Speech
\thanks{This work was supported by JST Moonshot Grant Number JPMJMS2011 and JST SPRING Grant Number JPMJSP2108.}
}

\author{
\authorblockN{Dong Yang, Yuki Saito, Wataru Nakata, and Hiroshi Saruwatari}
\authorblockA{The University of Tokyo, Japan \\
E-mail: ydqmkkx@gmail.com}
}


\maketitle
\pagestyle{empty}

\begin{abstract}
This paper introduces CraBERT, a pre-trained phoneme encoder (PPEnc) designed for efficient pre-training in text-to-speech (TTS). CraBERT employs a cascade-fusion architecture and a subword-phoneme alignment algorithm to integrate representations from a pre-trained subword-level BERT into a phoneme-level BERT. This design provides prior word- and sentence-level information, reducing the amount of pre-training required by the phoneme encoder. Subjective listening evaluations show that CraBERT achieves MOS values comparable to existing PPEncs after approximately one epoch of pre-training, whereas the baselines in our comparison are pre-trained for approximately ten epochs. These results demonstrate that CraBERT can efficiently learn representations suitable for improving the perceived naturalness and prosody of synthesized speech.
\end{abstract}


\section{Introduction}
In many text-to-speech (TTS) models~\cite{fastspeech2, gradtts, vits, vits2, styletts2}, the phoneme encoder serves as a critical component. It encodes phoneme tokens into phoneme representations and significantly enhances the quality and prosody of synthesized speech. BERT~\cite{bert}, characterized by its bi-directional attention architecture and masked language modeling (MLM) pre-training, has demonstrated a remarkable capacity to capture semantic features. This capability aligns well with the requirements of phoneme encoders in TTS models, leading to the development of several BERT-based pre-trained phoneme encoders (PPEncs).

PnG BERT~\cite{pngbert} pioneered concatenating phoneme and grapheme tokens as input with the MLM objective. While this strategy exploits phoneme-grapheme relationships, the joint input considerably increases the training and inference times. To address this limitation, subsequent PPEncs utilize only phoneme tokens as input and introduce several approaches to enhance the models' performance. Mixed-Phoneme BERT~\cite{mpbert} (MP BERT) addresses the lack of representation capacity due to the limited size of the phoneme vocabulary. It proposes BPE~\cite{bpe}-based sup-phoneme tokens to enhance the contextual representations, which are added with phoneme tokens as the input. Phoneme-Level BERT~\cite{plbert} (PL BERT) incorporates a phoneme-to-grapheme (P2G) prediction task during pre-training to introduce word-level semantic information. XPhoneBERT~\cite{xphonebert} uses vanilla BERT with the MLM objective to implement a multilingual PPEnc. When implementing MP BERT and PL BERT, we identified three characteristics of phonemes that notably impact phoneme-only PPEncs:
\begin{itemize}
    \item \textbf{Long sequence length}: During pre-training, PPEncs learn features at phoneme, word, and sentence levels. However, phoneme sequences are, on average, two to three times longer than their corresponding word sequences. This increased length renders phoneme-only PPEncs inefficient in learning word-level and sentence-level features.

    \item \textbf{Various phonetic notations}: Each language can have multiple phoneme types. Training multiple PPEncs for different phoneme types significantly increases computational resource consumption, highlighting the need for a more efficient pre-training method.
    
    \item \textbf{Limited vocabulary size}: As mentioned in MP BERT, phoneme vocabulary size is too small to provide sufficient representation capacity~\cite{vocab0, vocab1, vocab2}. Encoding semantics into a small set of phoneme tokens leads to semantic ambiguity and loss of details. Besides, the MLM task becomes less challenging, reducing its effectiveness as a pre-training objective.
\end{itemize}
Sup-phoneme tokens introduced in MP BERT address the issue of limited vocabulary size. However, the problem caused by long sequence length still hinders training efficiency.

Several works have attempted to enhance the phoneme encoder outputs of TTS models with subword-level BERT representations~\cite{cauliflow, parallelfusion}. These studies combine the two representations additively but do not pre-train the phoneme encoder. Moreover, existing approaches to the alignment of subwords and phonemes remain suboptimal. Since word-phoneme alignment can be obtained directly, Abbas et al.~\cite{cauliflow} average subword representations as word representations, which are then added with phoneme representations. This downsampling process leads to a loss of semantic and positional information. Wang et al.~\cite{parallelfusion} employ heuristic rules to align subword tokens and phoneme tokens and train an independent attention alignment module for inference. Rule-based alignment is complex and inaccurate, and they do not disclose the rules used. For the P2G objective in the pre-training of PL BERT, Li et al.~\cite{plbert} use a word-level tokenizer that avoids subword-phoneme alignment. However, word-level tokenizers can only tokenize out-of-vocabulary words as the ``unknown'' token, which limits the model's performance on rare words.

To overcome these obstacles, we propose \textbf{CraBERT}, a \textbf{C}ascade \textbf{R}epresentation \textbf{A}ligned pre-trained phoneme-level \textbf{BERT}. CraBERT leverages a pre-trained subword-level BERT to enhance representation capacity and provide prior word-level and sentence-level features. It then aligns and fuses these pre-trained subword representations with untrained phoneme representations via element-wise addition, enabling the phoneme-level BERT to learn TTS-relevant features with substantially less pre-training. In subjective evaluations, CraBERT trained for approximately one epoch achieved MOS values comparable to MP BERT and PL BERT trained for approximately ten epochs. Speech samples are available.\footnote{\url{https://ydqmkkx.github.io/CraBERT-Demo/}}

The primary contributions of this paper are:
\begin{itemize}
    \item CraBERT utilizes a pre-trained subword-level BERT to provide prior word- and sentence-level information, enabling efficient pre-training of its phoneme-level BERT.
    
    \item CraBERT achieves MOS values comparable to existing PPEncs using one-tenth as many pre-training steps. We also show that a masking rate higher than the conventional 15\% is beneficial under this efficient pre-training strategy.
\end{itemize}

As a secondary contribution, we introduce a data-driven subword-phoneme alignment algorithm based on dynamic time warping (DTW). It provides the necessary condition for fusing subword and phoneme representations within CraBERT.

\section{CraBERT}
\label{sec: crabert}

\subsection{Model}
\label{sec: crabert/model}

\begin{figure}[t]
  \centering
  \includegraphics[width=1.\linewidth]{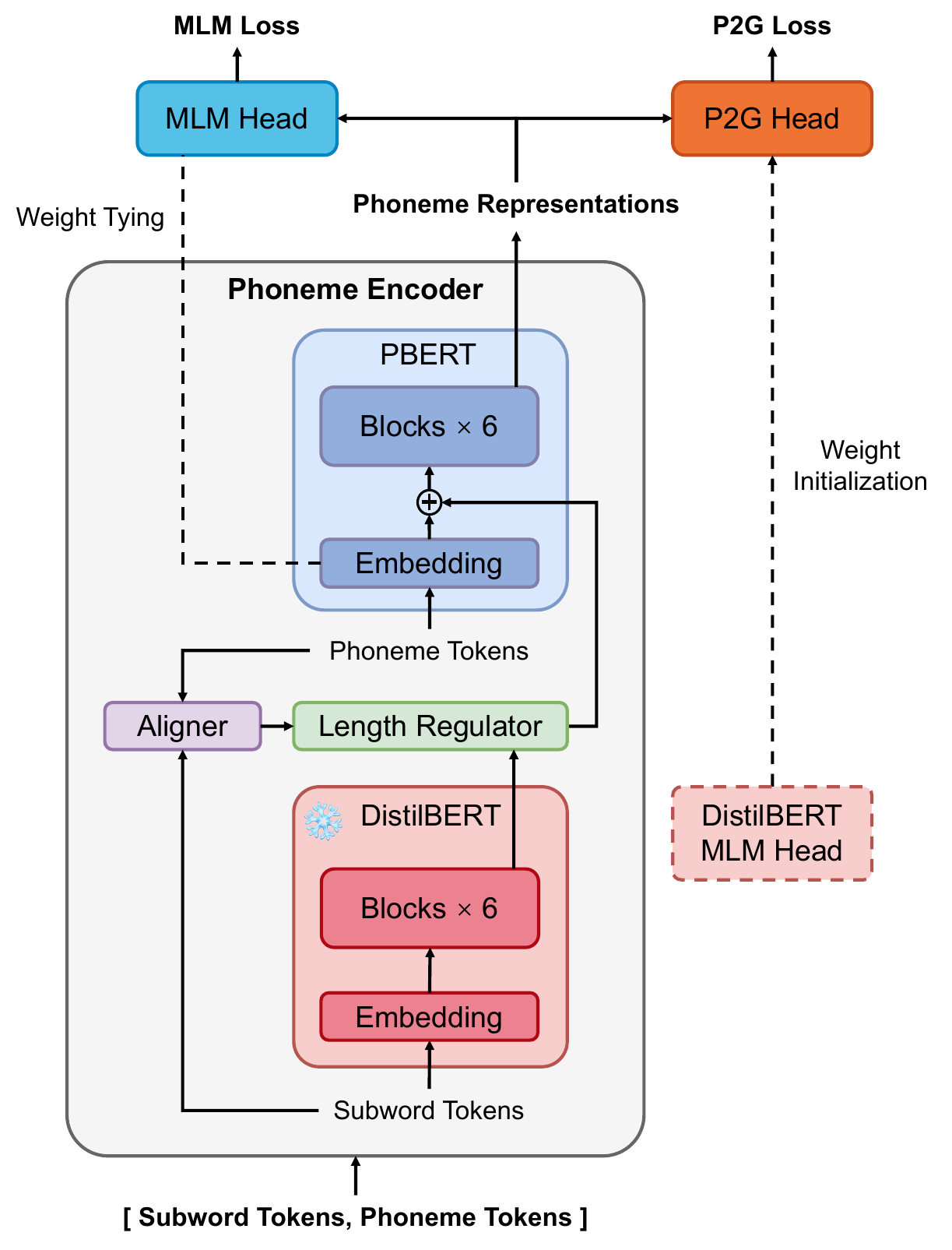}
  \caption{Model architecture and pre-training process.}
  \label{fig: model}
\end{figure}

The architecture and pre-training process of CraBERT are illustrated in Fig.~\ref{fig: model}. We employ a pre-trained DistilBERT\footnote{\url{https://huggingface.co/distilbert/distilbert-base-uncased}}~\cite{distilbert} to process input subword tokens and a vanilla BERT\textsubscript{BASE} to handle input phoneme tokens; the latter is referred to as \textbf{PBERT}. The aligner (Section~\ref{sec: CraBERT/aligner}) aligns input subword and phoneme tokens. Subsequently, the subword representations from DistilBERT are upsampled according to the alignment results via a length regulator module, which is similar to the one in FastSpeech 2~\cite{fastspeech2}. The aligned subword representations are then fused with the phoneme embeddings through element-wise addition before being fed into the BERT\textsubscript{BASE} blocks of PBERT, facilitating a cascade fusion of the two types of representations.

To ensure fair comparisons with our implementations of MP BERT and PL BERT (Section~\ref{sec: configurations/baseline}), we configure PBERT with relative positional encoding~\cite{relative} and 6 BERT\textsubscript{BASE} blocks, forming a 12-layer model in conjunction with DistilBERT.

\subsection{Pre-training}

During pre-training, DistilBERT is frozen to accelerate the training process and enhance stability. The pre-training tasks consist of MLM and P2G objectives, both utilizing cross-entropy loss:

\begin{itemize}
    \item \textbf{MLM}: We implement whole-word dynamic masking exclusively on phoneme tokens, adhering to the conventional masking strategy. Specifically, for the masked non-special phoneme tokens, 80\% are replaced with the \texttt{[MASK]} token, 10\% are replaced with random phoneme tokens, and 10\% remain unchanged. Additionally, for phoneme tokens replaced with the \texttt{[MASK]} token, we replace the corresponding subword representations with a trainable embedding.

    \item \textbf{P2G}: We have observed that applying P2G prediction to all phoneme tokens, as in PL BERT, adversely affects the training efficiency of the MLM objective. Therefore, the P2G objective only predicts the subword tokens corresponding to the masked phoneme tokens, similar to the prediction of masked sup-phoneme tokens in MP BERT. For simplicity, average pooling is not employed for loss calculation.
\end{itemize}

The two objectives can be formulated as follows, where $\theta$ denotes the module parameters, $\mathbf{p}$ and $\mathbf{g}$ represent the input phoneme and subword sequences respectively, and $M$ is the set of masked indices:
\begin{equation}
    \mathbf{r}_\mathbf{g} = \theta_{DistilBERT}\left( \mathbf{g} \right)
    \label{equation: r}
\end{equation}

\begin{equation}
    L_{MLM} = - \mathbb{E} \left[ \sum_{m \in M} \log P \left( p_m \mid \mathbf{p}, \mathbf{r}_\mathbf{g} ; \theta_{PBERT}, \theta_{MLM} \right) \right]
    \label{equation: mlm}
\end{equation}

\begin{equation}
    L_{P2G} = - \mathbb{E} \left[ \sum_{m \in M} \log P \left( g_m \mid \mathbf{p}, \mathbf{r}_\mathbf{g} ; \theta_{PBERT}, \theta_{P2G} \right) \right]
    \label{equation: p2g}
\end{equation}

Weight tying~\cite{tie} is a common strategy in BERT-based models to reduce parameters and enhance performance. We tie the weights between the embedding layer of PBERT and the output layer of the MLM head. Furthermore, we initialize the weights of the P2G head using the pre-trained MLM head of DistilBERT. Our preliminary experiments indicate that this initialization strategy accelerates the convergence of the P2G objective.

\subsection{Subword-phoneme aligner}
\label{sec: CraBERT/aligner}

Given the variability in word segmentation across different subword segmentation algorithms, we aim to develop a concise aligner applicable to all subword segmentation algorithms. Based on previous grapheme-to-phoneme (G2P) models, Futami et al.~\cite{phone-subword-align} proposed a two-step subword-phoneme alignment algorithm: first performing character-phoneme alignment, followed by subword-phoneme alignment. Character-phoneme alignment is a crucial step in training G2P models~\cite{align0, align1, align2}, typically using dictionaries or annotated speech corpora. However, in our task, we already have an existing G2P module that allows for the generation of extensive word-phoneme pairs from large textual corpora. Since Futami et al.~\cite{phone-subword-align} did not release their code and our own implementations performed worse, we follow their two-step method and propose a data-driven subword-phoneme aligner based on DTW:

\begin{itemize}
    \item \textbf{Training}: We define a probability matrix $\widetilde{\bm{\mathit{D}}} \in \mathbb{R}^{M \times N}$, where $\mathit{M}$ is the size of character set $\mathcal{C}$ and $\mathit{N}$ is the size of phoneme set $\mathcal{P}$. For a character index $\mathit{m}$, its corresponding character $\mathit{c}$ is obtained via $\mathit{\mathcal{C}(m)}=c$ and the inverse process is $\mathit{\mathcal{C}^{-1}(c)}=m$. Similarly, for the phoneme set, $\mathit{\mathcal{P}(n)}=p$ and $\mathit{\mathcal{P}^{-1}(p)}=n$. As outlined in Algorithm 1, we compute a value associated with the co-occurrence probability for each character-phoneme pair of a word. This value is based on the relative position distance $\mathit{d}$ of their corresponding sequences and decays as the distance increases, according to the decay factor function $\delta(d)$ in Equation~(\ref{equation: decay}):

    \begin{equation}
        \delta (d) = e^{-\alpha d^2}, \alpha=50
    \label{equation: decay}
    \end{equation}

    After training using BookCorpus~\cite{bookcorpus}, we obtain a distance matrix $\bm{\mathit{D}}\in \mathbb{R}^{M \times N}$.
    
    \item \textbf{Letter-phoneme alignment}: As depicted in Algorithm 2, in the forward stage, we utilize the pre-trained $\bm{\mathit{D}}$ to construct a cost matrix $\bm{\mathit{L}} \in \mathbb{R}^{I \times J}$ for the character and phoneme sequences of the word to be aligned. We employ a track matrix $\bm{\mathit{T}} \in (\mathbb{N}^2)^{I \times J}$ to record the locally optimal decisions, represented as directional indicators. In the backward stage, $\bm{\mathit{T}}$ enables efficient backtracking to reconstruct the optimal warping path.

    \item \textbf{Subword-phoneme alignment}: Because letter-subword alignment is directly obtainable, subword-phoneme alignment for each phoneme token is achieved by following the mapping sequence: phoneme $\rightarrow$ letter $\rightarrow$ subword.
    
\end{itemize}

\begin{algorithm}[t]
\caption{Distance Matrix Training}
\label{algorithm: training}
\begin{algorithmic}[1]
\State \textbf{Input:} training word set $\mathcal{W}$, character set $\mathcal{C}$, phoneme set $\mathcal{P}$, decay factor function $\delta$, $\epsilon=10^{-6}$
\State \textbf{Initialize:} probability matrix $\widetilde{\bm{\mathit{D}}} \gets \bm{\mathit{O}}_{M \times N}$

\For{each word in $\mathcal{W}$ with character sequence $\{c_{i}\}_{i=0}^{I-1}$ and phoneme sequence $\{p_{j}\}_{j=0}^{J-1}$}
    \For{$i=0$ \textbf{to} $I-1$}
        \For{$j=0$ \textbf{to} $J-1$}
            \State $\widetilde{\bm{\mathit{D}}}_{\mathcal{C}^{-1}(c_{i}), \mathcal{P}^{-1}(p_{j})} \gets \widetilde{\bm{\mathit{D}}}_{\mathcal{C}^{-1}(c_{i}), \mathcal{P}^{-1}(p_{j})} + \delta(\frac{i}{I} - \frac{j}{J})$
        \EndFor
    \EndFor
\EndFor

\For{each row $m \in [0, M-1]$ in $\widetilde{\bm{\mathit{D}}}$} \Comment{Normalization}
    \State 
    \[
        \widetilde{\bm{\mathit{D}}}_{m, \cdot} \gets \frac{\widetilde{\bm{\mathit{D}}}_{m, \cdot}}{\max(\max_{n \in [0, N-1]}\widetilde{\bm{\mathit{D}}}_{m, n}, \epsilon)}
    \]
\EndFor

\State \textbf{Output:} distance matrix $\bm{\mathit{D}} \gets \bm{\mathit{J}}_{M \times N} -\widetilde{\bm{\mathit{D}}}$
\end{algorithmic}
\end{algorithm}

\begin{algorithm}[t]
\caption{Character-Phoneme Alignment}
\label{algorithm: alignment}
\begin{algorithmic}[1]
\State \textbf{Input:} distance matrix $\bm{\mathit{D}}$, word $\mathit{w}$ with character sequence $\{c_{i}\}_{i=0}^{I-1}$ and phoneme sequence $\{p_{j}\}_{j=0}^{J-1}$
\State \textbf{Initialize:} 
    cost matrix $\bm{\mathit{L}} \gets \bm{\mathit{O}}_{I \times J}$, 
    track matrix $\bm{\mathit{T}} \gets \mathit{[(0, 0)]}_{I \times J}$, 
    alignment list $A \gets [\ ]$

\Procedure{Forward}{}
\For{$i=1$ \textbf{to} $I-1$}
    \State $\bm{\mathit{L}}_{i, 0} \gets \bm{\mathit{L}}_{i-1, 0} + \bm{\mathit{D}}_{\mathcal{C}^{-1}(c_{i}), \mathcal{P}^{-1}(p_{0})}$
    \State $\bm{\mathit{T}}_{i, 0} \gets (i-1, 0)$
\EndFor

\For{$j=1$ \textbf{to} $J-1$}
    \State $\bm{\mathit{L}}_{0, j} \gets \bm{\mathit{L}}_{0, j-1} + \bm{\mathit{D}}_{\mathcal{C}^{-1}(c_{0}), \mathcal{P}^{-1}(p_{j})}$
    \State $\bm{\mathit{T}}_{0, j} \gets (0, j-1)$
\EndFor

\For{$i=1$ \textbf{to} $I-1$}
    \For{$j=1$ \textbf{to} $J-1$}
        \State $t \gets \arg \min_{t \in \{
            (i-1, j), (i, j-1), (i-1, j-1)\}} \bm{\mathit{L}}_{t}$
        \State $\bm{\mathit{L}}_{i, j} \gets \bm{\mathit{L}}_{t} + \bm{\mathit{D}}_{\mathcal{C}^{-1}(c_i), \mathcal{P}^{-1}(p_j)}$
        \State $\bm{\mathit{T}}_{i, j} \gets t$
    \EndFor
\EndFor
\EndProcedure

\Procedure{Backward}{}
\State $(i, j) \gets (I-1, J-1)$
\State $A \gets [(c_i, p_j)]$
\While{$i>0$ \textbf{or} $j>0$}
    \State $(i,j) \gets \bm{T}_{i, j}$
    \State $A \gets A+[(c_i, p_j)]$
\EndWhile
\State Reverse the order of $\mathit{A}$
\EndProcedure

\State \textbf{Output:} alignment list $\mathit{A}$
\end{algorithmic}
\end{algorithm}

\section{Experimental configurations}
\label{sec: configurations}

\subsection{Baseline PPEncs}
\label{sec: configurations/baseline}

Among the latest PPEncs, we used MP BERT and PL BERT as our baseline models, given that XPhoneBERT focuses on multilingual capabilities without introducing architectural modifications. In their original papers, MP BERT and PL BERT diverged in BERT backbones, configurations, datasets, and the G2P modules. To facilitate more effective comparisons, we implemented them within a unified framework. We used a 12-layer vanilla BERT\textsubscript{BASE} as the backbone, with relative position encoding, 768 hidden size, and 12 self-attention heads. Specifically, for MP BERT, the BPE base dictionary of sup-phoneme was set as 30,000~\cite{mpbert}. For PL BERT's word-level tokenizer used in its P2G prediction objective, we opted for the Transformer-XL tokenizer\footnote{\url{https://huggingface.co/transfo-xl/transfo-xl-wt103}} that is consistent with PL BERT's official implementation\footnote{\url{https://github.com/yl4579/PL-BERT/blob/main/Configs/config.yml}}.

\subsection{Phoneme tokenizer}
\label{sec: configurations/tokenizer}

We developed a phoneme tokenizer based on the grapheme-to-phoneme (G2P) module from~\cite{g2p} due to its efficiency in G2P conversion, which was used to convert text sequences to ARPAbet phoneme sequences within the dataset. Unlike previous methods that separated phonemes of adjacent words with separators, we adopted the subword prefix of the WordPiece tokenizer~\cite{wordpiece}. For the phonemes of a word and any consecutive punctuation marks, we appended the prefix \#\# to all subsequent items after the first one. For example, the text \texttt{hello?!} is phoneme-tokenized as:
\begin{center}
\texttt{hh \#\#ah \#\#l \#\#ow ? \#\#!}
\end{center}

\subsection{Pre-training}
\label{sec: configurations/pretraining}

To ensure PPEncs receive a comparable amount of long-form textual information as subword-level language models, we limited the input phoneme sequences to a maximum length of 1,024. Our pre-training strategy and configuration were inspired by RoBERTa~\cite{roberta}, which demonstrated that large mini-batches can enhance optimization and achieve comparable or superior pre-training results. The pre-training configurations are as follows:
\begin{itemize}
    \item \textbf{Optimizer}: AdamW~\cite{adamw} with hyper-parameters ${\beta}_{1}=0.9$, ${\beta}_{2}=0.98$, ${\epsilon}=1 \times 10^{-6}$, and $L_{2}$ weight decay $\lambda= 0.01$.
    \item \textbf{Learning rate scheduler}: Linear scheduler, increasing from $0$ to $5 \times 10^{-4}$ during the initial $10\%$ warm-up training steps, then decreasing linearly to $0$.
    \item \textbf{Mini-batch size}: 2,000 sequences through gradient accumulation.
    \item \textbf{Hardware}: 8 $\times$ NVIDIA A100 40GB GPUs. 
    \item \textbf{Precision}: Mixed precision floating-point (FP16 and FP32).
\end{itemize}

In their respective papers, MP BERT was pre-trained on WMT news plus English Wikipedia for approximately 40 epochs, and PL BERT was pre-trained on English Wikipedia for about 10 epochs. To align with BERT and DistilBERT, we used BookCorpus plus English Wikipedia as our pre-training corpus. In our unified comparison, MP BERT and PL BERT were pre-trained for 90,000 steps, equivalent to approximately 10 epochs, whereas CraBERT-1e was pre-trained for only 9,000 steps, or approximately one epoch.

\subsection{TTS}
We evaluated the performance of PPEncs through the downstream TTS task. To better demonstrate the impact of various PPEncs on the prosody and naturalness of synthetic speech, we conducted TTS experiments using a multi-speaker dataset. It was derived from the ``train-clean-360'' subset of the LibriTTS-R corpus~\cite{librittsr} and annotated with respiratory pauses~\cite{yang23pause} and breath sounds~\cite{yang24breath}. We selected VITS~\cite{vits} as the backbone TTS model, replacing its phoneme encoder with different PPEncs. The TTS models were trained for 250 epochs with a mini-batch size of 80 on a single A100 GPU. We employed the AdamW optimizer with PyTorch's default hyper-parameters and the same scheduler as detailed in Section~\ref{sec: configurations/pretraining}, with a peak learning rate of $3 \times 10^{-4}$. During inference, we randomly sampled 50 speaker-text pairs from the test set to generate speech for subjective evaluations.


\subsection{Subjective evaluation}
Because perceived prosody and naturalness are not fully captured by conventional objective metrics, we focus on subjective listening evaluations, following prior PPEnc studies~\cite{mpbert, plbert}. We conducted mean opinion score (MOS) tests on Prolific\footnote{\url{https://www.prolific.com/}} to evaluate the synthesized speech. Native English listeners rated the naturalness and prosody of each synthesized speech sample on a five-point scale (1 = bad, 2 = poor, 3 = fair, 4 = good, and 5 = excellent).

\section{Experimental results and discussion}
\label{sec: experiments}

\subsection{Masking rate for efficient pre-training}
\label{sec: experiments/mask}

\begin{table}[t]
  \caption{MOS results for different masking rates. CI denotes the 95\% confidence interval.}
  \label{tab: masking results}
  \centering
  \small
  \setlength{\tabcolsep}{12pt}
  \begin{tabular}{cc}
    \toprule
    \textbf{Masking rate} & \textbf{MOS $\uparrow$ $\pm$ CI}\\
    \midrule
    15\% & 3.10 $\pm$ 0.16\\
    30\% & 3.23 $\pm$ 0.18\\
    45\% & 3.25 $\pm$ 0.17\\
    60\% & 3.25 $\pm$ 0.17\\
    75\% & \textbf{3.29 $\pm$ 0.17}\\
    90\% & 3.15 $\pm$ 0.19\\
    \bottomrule
  \end{tabular}
\end{table}

In our initial experiments, we observed that applying a 15\% masking rate for CraBERT's pre-training, following the conventional masking strategy of BERT-like language models, did not yield satisfactory results. As shown in equations~\ref{equation: mlm} and~\ref{equation: p2g}, the subword representations generated by DistilBERT, which are inherently rich in contextual information, significantly simplify the predictions for both objectives when served as prior conditions. This led us to hypothesize that the 15\% masking rate might be insufficient for CraBERT, and its pre-training could benefit from a higher masking rate.

To explore this hypothesis, we conducted a series of pre-training experiments on CraBERT, gradually increasing the masking rate from 15\% to 90\% in 15\% increments. All models were pre-trained for 9,000 steps (about 1 epoch) and subsequently fine-tuned on VITS to generate synthetic speech. We recruited 50 native English listeners in the MOS test, each of whom rated 18 speech samples. As presented in Table~\ref{tab: masking results}, the MOS generally increases with the masking rate, reaches its highest observed value at 75\%, and then decreases at 90\%. This result suggests that CraBERT benefits from a higher masking rate than the conventional 15\%. We therefore selected 75\% for the subsequent experiments.

\subsection{Comparisons among phoneme encoders}
\label{sec: experiments/comparisons}

\begin{table}[t]
  \caption{MOS results across phoneme encoders. CI denotes the 95\% confidence interval.}
  \label{tab: comparative results}
  \centering
  \small
  \setlength{\tabcolsep}{5pt}
  \begin{tabular}{lrc}
    \toprule
    \textbf{Encoder} & \textbf{Pre-training steps} & \textbf{MOS $\uparrow$ $\pm$ CI}\\
    \midrule
    Ground Truth & -- & 3.60 $\pm$ 0.18\\
    \midrule
    Baseline & 0 & 2.83 $\pm$ 0.18\\
    MP BERT & 90,000 & 3.14 $\pm$ 0.17\\
    PL BERT & 90,000 & 3.13 $\pm$ 0.19\\
    \addlinespace[0.4em]
    CraBERT\textsubscript{para}-0e & 0 & 2.90 $\pm$ 0.18\\
    CraBERT-0e & 0 & 3.09 $\pm$ 0.17\\
    CraBERT-1e & 9,000 & 3.21 $\pm$ 0.18\\
    CraBERT-10e & 90,000 & 3.15 $\pm$ 0.19\\
    \bottomrule
  \end{tabular}
\end{table}

We refer to the CraBERT model in Section~\ref{sec: experiments/mask}, pre-trained for 9,000 steps with a 75\% masking rate, as \textbf{CraBERT-1e}. Additionally, we pre-trained CraBERT with the same 75\% masking rate for 90,000 steps, denoting it as \textbf{CraBERT-10e}. To evaluate our proposed methods, we conducted comparative experiments with three baseline methods:
\begin{itemize}
    \item \textbf{Baseline VITS}: For fair comparisons, we replaced the phoneme encoder in VITS with an untrained 12-layer PBERT.
    
    \item \textbf{Baseline PPEncs}: MP BERT and PL BERT.
    
    \item \textbf{Baseline fusion approach}: We explored the parallel fusion approach proposed in~\cite{cauliflow, parallelfusion}, which fuses BERT representations with phoneme encoder outputs. Since our pre-training strategy cannot be applied to this approach, we added the outputs of DistilBERT and PBERT without pre-training to construct \textbf{CraBERT\textsubscript{para}-0e}, which was assessed alongside \textbf{CraBERT-0e}.
\end{itemize}

We recruited 50 native English listeners in the MOS test, each of whom rated 24 speech samples. The results are shown in Table~\ref{tab: comparative results}. Since the masking-rate and model-comparison experiments used separate listening tests, MOS values should be compared within each table rather than across Tables~\ref{tab: masking results} and~\ref{tab: comparative results}.

\begin{itemize}
    \item All phoneme encoders with either partially or fully pre-trained parameters demonstrated improvements in perceived naturalness compared to the untrained phoneme encoder in baseline VITS. 

    \item CraBERT-1e, trained for 9,000 steps, obtained a MOS comparable to MP BERT and PL BERT trained for 90,000 steps. Thus, CraBERT used one-tenth as many pre-training steps to reach comparable perceived naturalness and prosody.

    \item CraBERT-10e did not provide a further MOS improvement over CraBERT-1e, suggesting that prolonged phoneme-level pre-training is unnecessary for the evaluated TTS setting.
    
    \item Regarding fusion approaches, CraBERT-0e with cascade fusion achieved a higher MOS than CraBERT\textsubscript{para}-0e with parallel fusion, highlighting the efficacy of cascade fusion.
\end{itemize}

The comparisons among CraBERT-0e, CraBERT-1e, and CraBERT-10e indicate that CraBERT reaches its useful representation quality after only moderate pre-training of PBERT. We hypothesize that this efficiency stems from the word- and sentence-level information already supplied by DistilBERT, allowing PBERT to focus on phoneme-level features relevant to TTS. The absence of further improvement from CraBERT-1e to CraBERT-10e may also reflect a discrepancy between phoneme-level contextual features acquired through language modeling and those required for TTS. We leave further analysis of this discrepancy to future work.

\subsection{Pre-training efficiency}

On 8 $\times$ A100 GPUs, one epoch of pre-training requires about 19, 27, and 29 hours for CraBERT, MP BERT, and PL BERT, respectively. CraBERT's lower per-epoch cost is attributed to freezing DistilBERT. More importantly, CraBERT-1e reaches a comparable MOS after about 19 hours. Based on the measured per-epoch times, the 10-epoch MP BERT and PL BERT baselines require approximately 270 and 290 hours, corresponding to pre-training speedups of about $14\times$ and $15\times$, respectively. Their inference speeds are comparable on a single A100 GPU.

\section{Conclusion}
In this study, we proposed CraBERT, a PPEnc designed to reduce the pre-training required for TTS. By aligning and fusing representations from a pre-trained subword-level BERT, CraBERT provides word- and sentence-level information before phoneme-level pre-training. Subjective evaluations showed that CraBERT trained for approximately one epoch achieved MOS values comparable to PPEnc baselines trained for approximately ten epochs. Under the same setup of 8 A100 GPUs, CraBERT used one-tenth as many training steps and achieved estimated pre-training speedups of about $14\times$ and $15\times$ over MP BERT and PL BERT, respectively. Future work will focus on determining the optimal degree of pre-training and exploring the distinctions between phoneme features in language models and TTS models.

\bibliographystyle{IEEEbib}
\bibliography{refs}
\end{document}